\documentclass{aa}

\newcommand{\be}{\begin{equation}}
\newcommand{\ee}{\end{equation}}
\newcommand{\bdm}{\begin{displaymath}}
\newcommand{\edm}{\end{displaymath}}

\input psfig.tex
\def\lsim{\lower.5ex\hbox{$\; \buildrel < \over \sim \;$}}
\def\gsim{\lower.5ex\hbox{$\; \buildrel > \over \sim \;$}}
\usepackage{graphicx}

\begin{document}
\onecolumn

\input psfig.tex
\def\lsim{\lower.5ex\hbox{$\; \buildrel < \over \sim \;$}}
\def\gsim{\lower.5ex\hbox{$\; \buildrel > \over \sim \;$}}

\title{Accretion powered spherical wind in general relativity}
\author{Tapas K. Das\inst{1,2}}
\offprints{Tapas K. Das \\ \email{tapas@iucaa.ernet.in}}

\institute{Astronomy Unit, Queen Mary \& Westfield College, Mile End Rd, London E1 4NS, UK\\
\email{T.Das@qmw.ac.uk}
\and
Inter University
 Centre For Astronomy And Astrophysics, Post Bag 4 Ganeshkhind, Pune 411 007,
India\\
\email{tapas@iucaa.ernet.in}}
\date{Received: 9$^{th}$ May, 2001/ Accepted: 18$^{th}$ June, 2001 }
\authorrunning{Das T. K.}
\titlerunning{Accretion Powered Spherical Wind in General Relativity}
\abstract{Using full general relativistic calculations,
we investigate the possibility of generation of mass outflow from spherical
accretion onto non-rotating black holes. Introducing a relativistic
hadronic-pressure-supported steady, standing, spherically-symmetric shock surface
around a Schwarzschild black hole as the effective physical barrier that
may be responsible for the generation of spherical wind, we calculate the mass
outflow rate $R_{\dot m}$ in terms of three accretion parameters and one outflow
parameter by simultaneously solving the set of general relativistic hydrodynamic
equations describing spherically symmetric, transonic, polytropic
accretion and wind around a Schwarzschild black hole. Not only do we provide a
sufficiently plausible estimation of $R_{\dot m}$, we also 
successfully
study the dependence and variation of this rate on various physical parameters
governing the flow. Our calculation indicates that independent of
initial boundary conditions, the baryonic matter content of this
shock-generated wind always correlates with post-shock flow temperature.
}

\maketitle
\hrule
\begin{center}
{\bf Published in Astronomy and Astrophysics, 2001, v.376, p.697-707.}
\end{center}
\hrule

\section {Introduction}
\noindent
Accretion onto  massive celestial bodies is believed to be responsible
for energy generation mechanism in many astrophysical objects including
galactic binary systems and extra-galactic objects like Quasars and
AGNs. 
Investigation of transonic,
spherically symmetric accretion onto gravitating astrophysical
objects was carried out from a purely Newtonian approach as well as 
using a pseudo-general relativistic and 
complete general relativistic framework
(Hoyle \& Littleton 1939, Bondi 1952, Michel 1972, Shapiro 1973a,b,
Blumenthal \& Mathews 1976, Begelman 1978, Brinkmann 1980, Malec 1999, 
Das 2001, Das \& Sarkar 2001, Sarkar \& Das 2001).
However, in works mentioned above, spherical accretion was shown to 
exhibit essentially directed motion (radial inflow) 
which smoothly
passes through a sonic point and no discontinuity in accretion 
onto a black hole was taken into account.
Nevertheless, it was understood that accretion onto a black hole could be
efficient in producing radiative energy if the directed in-fall motion could be
randomized close to the hole by some suitable mechanism the best of which
would be shock formation by some means. For radial accretion, a novel mechanism
has been proposed (Protheros \& Kazanas 1983, Kazanas \& Ellision 1986,
hereafter PK83 and KE86 respectively)
in which the kinetic energy of spherically
accreting material has been randomized by proposing a steady, collision-less,
relativistic hadronic-pressure-supported spherically symmetric 
shock around a Schwarzschild black hole which produces a nonthermal spectrum
of high energy (relativistic) protons.
A thermal distribution of particles may also be produced at the shock
which radiates by thermal bremstrahlung but only the radiation
resulting from the non-thermal (which might actually contain most of
the energy, see Axford 1981) was considered in their model.
Since radial in-fall (freely falling situation) in a gravitational potential is
unlikely to release amounts of energy
sufficient to produce the observed high luminosity and high energy tail of the
spectrum for most of the quasars, it has been suggested in PK83 and KE86
that the relativistic particles required to account for the observed radio,
X-ray and $\gamma$-ray emission may result from shock acceleration
of a part of in-falling plasma via first order Fermi acceleration
which transforms directed kinetic energy into 
relativistic particle
energy and thus produces relativistic protons with high efficiency. A 
steady-state situation is assumed to have developed where a standing collision-less
spherically symmetric shock is proposed to form.\\
\noindent
The question of the existence of a steady shock for spherical accretion
deserve some elaborate discussion and clarification.
As the absence of considerable angular momentum in accreting material does not supports 
standard  Rankine-Hugoniot type
shock formation, assumption of a steady and standing
spherically-symmetric shock for freely falling material, may apparently
contradict the concept of shock formation in general fluid flow.
The situation, however, is different if the
relativistic protons of sufficient energy density are present around the
black hole. It is worth studying
what happens when freely falling supersonic
(supersonic, because, as we will see later, spherical accretion
with a reasonably realistic
value of 
inflow energy and accretion rate 
crosses its sonic point far away, $\gsim$ hundreds of
Schwarzschild radius or even more, from the black hole) plasma approaches close
(some factor of tens of Schwarzschild radius or even less) to
the black hole. Due to various possible instabilities in supersonic plasma
flow, any small disruption mechanism may slow down the flow and create
a `piston' which produces a shock. Thus the initial formation of
shocks may be attributed either to convection of ambient relativistic particles
or to dissipation in magnetic field as suggested in the literature (see PK83, KE86,
and references therein). Once the shock is produced, it accelerates a part
of the in-falling matter to relativistic energies. Infalling thermal particles
with free fall velocity are then assumed to be shock-accelerated via first
order Fermi acceleration 
(Axfrod, Leer \& Skardon 1977, Blandford \& Ostriker 1978,
Blandford \& Eichler 1987, and references therein, Kirk \& 
Duffy, 1999, and references therein) 
and relativistic protons will be produced. 
These shock accelerated 
relativistic protons were assumed to be trapped within a radius comparable to
the radius of the shock surface by some mechanism (possibly by 
small amount of stochastic magnetic fields tied
to the accreting matter, which is small enough to
make the hydrodynamical approximation 
valid to describe the flow) because had it been the case that those protons would not
be confined to the central region of the quasars, spallation of nuclei in the
outer gas cloud responsible for the line emission in Quasars would result in an
overabundance of boron (Baldwin et. al. 1977), contrary to the observation. In
addition, the matter density in this region may be assumed to be sufficiently
high so that the relativistic protons are depleted mainly by nuclear interactions
which, possibly could set a lower limit (though not very restrictive) on the
accretion rate for a given black hole mass (PK83).
\noindent
It was also suggested in PK83 that if the particles accelerated at the shock were
neucleons rather than electrons, then, due to their negligible Compton losses,
the Compton catastrophe
could be avoided.\\
\noindent
Because of the fact that
the kinetic energy of these relativistic particles are much larger than
their gravitational potential energy,
and because their scattering mean free path
could be larger than the black hole radius,
\footnote{Furthermore, the converging magnetic field, if present, is also
expected to mirror the relativistic particles and help prevent them from being
absorbed by the black hole, see KE86.}
these relativistic protons suffer practically no Compton losses and are not
instantly swallowed by the black hole, rather they
scatter several times and undergo
multiple inelastic nuclear collisions and 
produce pions (${\pi}^{\pm,0}$)
before being captured by the black hole (PK83, KE86).
Pions generated by this process decay into relativistic electrons,
positrons,
neutrinos and antineutrinos and produce high energy $\gamma$-rays
(see \S 2 of Das (1999) and references therein for various particle 
production and decay mechanism).
As the relativistic protons are assumed to lose energy only through the above
mentioned nuclear collisions and since the proton-proton collision time
scale ${\tau}_{pp}$ would, in general, be much larger compared to the free-fall 
time scale, a sufficiently high density of relativistic protons
could be maintained which provides radially outward pressure sufficient
enough to support the sustenance of a steady and
standing shock. The shock is, thus,
self-supported.\\
\noindent
In a previous work (Das 1999), we have
explicitly calculated the exact location (radial
distance measured from the 
central accretor in units of Schwarzschild radius $r_g=\frac{2GM_{BH}}{c^2}$,
where $M_{BH}$ is the mass of the black hole, $G$ is the Universal gravitational 
constant and $c$ is the velocity of light in vacuum)
of the above mentioned steady, standing, spherically-symmetric shock
in terms of only three accretion parameters, namely, the specific energy of 
the flow, accretion rate (scaled in units of Eddington rate) and the adiabatic
index of the flow for spherically-symmetric, transonic accretion of
adiabatic fluid onto a Schwarzschild black hole. 
By solving the set of hydrodynamic equations describing the 
motion of accreting material under the influence of pseudo-Schwarzschild potential
proposed by Paczy\'nski \& Wiita (1980), we have shown there 
that it is possible to construct a self-consistent inflow-outflow system
where a part of the accreting material may be blown as spherical wind from the
spherical shock surface and the mass outflow rate $R_{\dot m}$ (the measure of the 
fraction of accreting material being `kicked off' as wind) was computed 
in terms of various accretion as well as shock parameters. However,
calculation presented in our previous work was based on the valid assumption that 
spherically symmetric, transonic flow onto a Schwarzschild black hole 
can be well approximated using a suitable modified Newtonian potential 
and no general relativistic 
calculations were performed to study the flow profile.\\
\noindent
In this paper, we would like to construct a self-consistent, 
spherically symmetric, polytropic, transonic, non-magnetized accretion-wind
system using a full general relativistic description of flow structure around a 
Schwarzschild
 black hole. In doing so, we would like to perform the following steps:\\
\noindent
(a) We will write down the set of general relativistic
hydrodynamic equations of motions describing the 
accretion and wind in Schwarzschild metric.\\
\noindent
(b) Incorporating a suitable equation of state, we will solve the equations to get 
a general flow profile for accreting material where any dynamical (flow velocity,
Mach number of the flow etc.) as well as thermodynamic (flow temperature, density,
pressure etc.) quantity
can be explicitly calculated at any radial distance
(measured from the accretor in units of $r_g$) in terms of only three
accretion parameters, namely, the total specific
energy and adiabatic index of the flow and
accretion rate scaled in units of Eddington rate.\\
\noindent
(c) We will use the formula for shock location obtained in Das (1999) with 
the assumption that direct use of that formula in this paper would not be 
quite unjustified because procedure to calculate the shock location in 
modified Newtonian description would
roughly be unchanged also in the full general relativistic framework. 
This is because the shock location is obtained by equating the 
two alternative expressions for luminosities of the same object, hence
any correction term arising from the red-shift (gravitational or cosmological)
would cancel out.
Using the values of dynamical (and thermodynamic) variables
computed in general relativistic framework in terms of three accretion parameters 
mentioned in (b), we will calculate the shock standoff distance.
As the condition necessary
for the development and sustenance of such a shock is satisfied when
at shock location, the Mach number of the inflow $M_{sh}$ is considerably high
(high supersonic flow at the time
it encounters the shock, Ellison \& Eichler 1984), we concentrate
on polytropic accretion with such a set of values of conserved
specific energy ${\cal E}$
and accretion rate (scaled in the unit of Eddington rate)
${\dot M}_{Edd}$ which produces a high shock Mach number solution.\\
\noindent
(d) We will take this hadronic pressure-supported shock surface to be 
the generating surface of mass outflow.
At the shock surface,
density of the post-shock material as well as the post-shock
flow temperature shoots up and dynamical flow velocity falls
down abruptly.
In other words, highly supersonic inflow becomes
subsonic and accreting matter becomes shock-compressed at this surface.
Matter starts getting hotter and denser and
starts piling up on the shock surface. The post-shock relativistic hadronic pressure
and thermal pressure (pressure generated by high temperature
produced at the shock)
then gives a kick to the piled-up
matter, the result of which is the ejection of outflow from the shock surface.
Thus a hot and dense spherical shock surface serves as the
`effective' physical atmosphere regarding the generation of mass
outflow from matter accreting onto black holes.
At the shock, entropy is generated
and the outflow as well as post-shock inflow will have higher
entropy density for the same specific energy.
For this type of inflow, accretion is expected to proceed smoothly
after a shock transition, since successful subsonic solutions
have been constructed for accretion onto black holes
embedded within normal stars with the boundary condition
${u = c}$ (Flammang 1982); where $u$ is the in-fall velocity of matter and $c$ is the
velocity of light in vacuum.\\
\noindent
It is to be mentioned here that one fundamental criterion
for formation of hydrodynamic outflows is that
the outflowing wind should have a positive Bernoulli constant which means
that the matter in the post-shock region is able to escape
to infinity. However, positiveness in Bernoulli's constant may lead to another
situation as well where shock may quasi-periodically originate
at some certain radius and propagate outwards without formation of
outflows. So formation of outflows is one of the possible scenarios
when we focus on the positive energy solutions.
In this paper we concentrate
only on solutions producing outflows, thus we 
use only the positive Bernoulli constant
throughout our model which is standard  practice for studying
hydrodynamic winds. 
Another assumption made in this paper 
is to treat the accreting as well as post-shock matter as a single temperature
fluid, the temperature of which is basically characterized by proton temperature.\\
\noindent
The plan of this paper is as follows. In next section we describe how to formulate
and solve the governing equations. In \S 3, we present our results. In \S 4, the
results will be reviewed and conclusions will be drawn.
\section{Governing equations and solution procedure}
\subsection {Inflow model}
We assume that a Schwarzschild type black hole
spherically accretes fluid
obeying  a polytropic equation of state. The density of the fluid is $\rho(r)$,
$r$ being the radial distance measured in the
unit of Schwarzschild radius $r_g$. We also assume that the accretion rate
(in the unit of
Eddington rate ${\dot M}_{Edd}$) is not a function
of $r$ and we ignore the
self-gravity of the flow.
For simplicity of calculation, we choose the geometric unit 
where the unit of length is scaled in units of $r_g$, units of velocity
in units of $c$ . All other physically relevant 
quantities can be expressed likewise. We also set $ G=c=1$ in the system of 
units used here.
\\
\noindent
For a Schwarzschild metric of the form 
$$
ds^2=dt^2\left(1-\frac{1}{r}\right)-dr^2{\left(1-\frac{1}{r}\right)}^{-1}-r^2{\left(d{\theta}^2+sin^2{\theta}{d\phi^2}\right)}
$$
the energy momentum tensor $T^{{\mu}{\nu}}$ for a perfect fluid
can be written as
$$
T^{\mu\nu}={\epsilon}u^{\mu}u^{\nu}+p\left(u^{\mu}u^{\nu}-
g^{\mu\nu}\right)
$$
where ${\epsilon}$ and $p$ are proper energy density and pressure of the 
fluid (evaluated in the local inertial rest frame of the fluid)
respectively and $u^{\mu}$ is the four velocity commonly known as
$$
u^{\mu}=\frac{dx^{\mu}}{ds}
$$
Equations of motion which are to be solved for our purpose are,\\
1) Conservation of mass flux or baryon number conservation:
$$
{\left({\rho}{u_{\mu}}\right)}_{;~{\mu}}=0
\eqno{(3a)}
$$
and \\
2) Conservation of momentum or energy flux (general relativistic Euler 
equation obtained by taking the four divergence of $T^{{\mu}{\nu}}$):
$$
\left({\epsilon}+p\right){u_{{\mu};{\nu}}}u^{\nu}=
-p_{,\mu}-u_{\mu}p_{,\nu}u^\nu_{,}
\eqno{(3b)}
$$
where the semicolons denote the covariant derivatives. \\
\noindent
Following Michel (1972), one can rewrite Eq. 3(a) and Eq. 3(b) for 
spherically symmetric flow as 
$$
4\pi{\rho}ur^2={\dot M}_{in}
\eqno{(4a)}
$$
and
$$
\left(\frac{p+\epsilon}{\rho}\right)^2\left(1-\frac{1}{r}+u^2\right)={\bf C}
\eqno{(4b)}
$$
as two fundamental conservation equations for time-independent 
hydrodynamical flow of matter on to a
Schwarzschild black hole without back-reaction of the
flow on to the metric itself. ${\dot M}_{in}$   being the mass
accretion rate and ${\bf C}$ is some constant 
(related to the total enthalpy influx)
to be evaluated
for a specific equation of state.\\
\noindent
For polytropic equation of state i.e,
$$
p=K{\rho}^{\gamma_{in}}
$$
(where $\gamma_{in}$ is the adiabatic index of the inflow)
and defining ${\dot {\cal M}}$
(where ${\dot {\cal M}}=
{\dot M}_{in}{\gamma_{in}}^{\frac{1}{\gamma_{in}-1}}
K^{\frac{1}{\gamma_{in}-1}}$ is a measure of the 
specific entropy of the flow)
to be another
conserved quantity of the flow
except at the shock location (because shock generates entropy), one can rewrite
conservation equations 4(a) and 4(b) as (Das \& Sarkar 2001, and
references therein):
$$
{\cal E}=hu_t=\left(\frac{p+\epsilon}{\rho}\right)
{\left(\frac{1-\frac{1}{r}}{1-u^2}\right)}^{\frac{1}{2}}
\eqno{(5a)}
$$
and
$$
{\dot {\cal M}}=4{\pi}\left[\frac{\left(\gamma_{in}-1\right)a^2}{\gamma_{in}
-\left(a^2+1\right)}\right]^{\left(\frac{1}{\gamma_{in}-1}\right)}
uu_tr^2
\eqno{(5b)}
$$
where ${\cal E}, h,u_t$  and $a$ are the total
conserved specific energy( which include the rest mass of matter),
specific enthalpy $\left(h=\left[\frac{\gamma_{in}-1}{\gamma_{in}-\left(a^2+1\right)}\right]\right)$,
specific binding
energy and local adiabatic sound speed
respectively. Eqn. 5(b) may be considered as the outcome of
the conservation
of mass and entropy along the flow line.
The expression for the adiabatic sound speed
$a$ can be written as (Weinberg 1972, Frank et. al.
1992):
$$
a={\left(\frac{{\partial}p}{{\partial}\epsilon}\right)}^{\frac{1}{2}}
_{\cal S}=\sqrt{\frac{{\gamma_{in}}{p}}{\rho}}=
\sqrt{\frac{{\gamma_{in}}{\kappa}{T}}{{\mu}{m_H}}}
\eqno{(6)}
$$
where $T$ is the flow temperature, $\mu$ is the mean molecular weight and
$m_H{\sim}m_p$ is the mass of the hydrogen atom
and $\kappa$ is Boltzmann's constant. The subscript ${\cal S}$ indicates that 
differentiation is performed at constant specific entropy.\\
\noindent
One can now easily derive the expression for velocity
gradient $\left(\frac{du}{dr}\right)$ (by differentiating Eq. 5(a) and 5(b))
as 
$$
\frac{du}{dr}=\frac{u\left(1-u^2\right)\left[a^2\left(4r-3\right)-1\right]}
{2r\left(r-1\right)\left(u^2-a^2\right)}
\eqno{(7a)}
$$
Since the flow is assumed to be smooth everywhere, if
the denominator of Eq. 7(a)  vanishes at any radial distance
$r$, the numerator must also vanish there to maintain the
continuity of the flow. One therefore arrives at the so
called `sonic point (alternately, the `critical point')
conditions'  by simultaneously making
numerator and denominator of Eq. 7(a) equal to zero and
the sonic point conditions can be expressed as follows
$$
u_c=a_c=\sqrt{\frac{1}{4r_c-3}}
\eqno{(7b)}
$$
For a specific value of ${\cal E}$ and $\gamma_{in}$,
location of sonic point $r_c$
can be obtained by solving the following equation
algebraically
$$
64r_c^3\left({\cal E}^2-1\right)+
16r_c^2\left(2{\cal E}^2{\varphi_{in}^2}-9\right)
+4r_c\left({\cal E}^2{\varphi_{in}^2}-27\right)+27=0
\eqno{(7c)}
$$
where ${\varphi_{in}}=\left(\frac{3\gamma_{in}+2}{\gamma_{in}-1}\right)$. \\
\noindent
To
determine the behaviour of the solution near the sonic
point, one needs to evaluate the value of $\left(\frac{du}{dr}\right)$
at that
point (we denote it by $\left(\frac{du}{dr}\right)_c$)
by applying L 'Hospitals' rule on Eq. 7(a). It is
easy to show that $\left(\frac{du}{dr}\right)_c$
can be obtained by solving the
following quadratic equations algebraically:
$$
\left(\frac{du}{dr}\right)^2_c+\frac{\left(\gamma_{in}-1\right)
\left(16r^2_c-16r_c-8{\gamma_{in}}r_c+6\gamma_{in}+3\right)}
{3r_c\left(4r_c-3\right)^{\frac{3}{2}}}
\left(\frac{du}{dr}\right)_c
$$
$$
+
\frac{\left(\gamma_{in}-1\right)\left(2r_c-1\right)\left(24r^2_c-28r_c-8r^2_c\gamma_{in}
+4r_c\gamma_{in}+3\gamma_{in}+6\right)}
{2r_c\left(4r_c-3\right)^{\frac{5}{2}}\left(r_c-1\right)}
=0
\eqno{(7d)}
$$
It is now quite straightforward to simultaneously
solve Eq. 5(a) and Eq. 5(b) to
get the integral curves of the flow 
(curves showing variation of Mach number of the flow with radial distance)
for a fixed value
of ${\cal E}$ and $\gamma_{in}$. Detail methodology for this purpose will 
be discussed
in \S 2.3.\\
In this work we normally use the value of $\gamma_{in}$
to be $\frac{4}{3}$. Though far away from
the black hole, optically thin accreting plasma
may not be treated as 
ultra-relativistic fluid 
(by the term `ultra-relativistic' and `purely non-relativistic'
we mean a flow with 
$\gamma=\frac{4}{3}$ and $\gamma=\frac{5}{3}$ respectively,
according to the terminology used in 
Frank et. al. 1992) in general, close to the
hole it {\it always} advects with enormously large radial velocity and
could be well-approximated as ultra-relativistic flow.
As because our main region of interest,
the shock formation zone, {\it always} lies close to the black hole
(a few tens of $r_g$ away from the hole
or sometimes even less, see results and figures in \S 3),
we believe that
it is fairly justifiable to assign the value $\frac{4}{3}$ for
$\gamma_{in}$ in our work.
However, to rigorously model a real flow without any assumption,
a variable polytropic index
having proper functional dependence on radial distance
(i.e., $\gamma~\equiv~\gamma(r)$) might be considered
instead of using a constant $\gamma_{in}$,
and equations of
motion might be formulated accordingly, which we did not attempt
here for the sake of simplicity.
Nevertheless, we keep our option open for values of $\gamma_{in}$
other than $\frac{4}{3}$ as well and investigated the outflow
for an range of values of $\gamma_{in}$ for a specific
value of ${\cal E}$ and ${\dot M}_{Edd}$ (Fig. 5a, \S 3.3.3).
The same kind of investigations could be performed for a variety of values of
${\cal E}$ and ${\dot M}_{Edd}$ and a set of results may be obtained
with  a wide range of values of $\gamma_{in}$ which tells that our
calculation is not restricted to the value $\gamma_{in}~=~\frac{4}{3}$
only; rather the model is general enough to deal with all possible
value of $\gamma_{in}$ for polytropic accretion.
\subsection{Shock parameters and the outflow model}
\noindent
As already mentioned, a steady, collision-less shock forms
due to the instabilities in the plasma flow.
We assume that for our model,
the effective thickness of the shock $\Delta_{sh}$ is small enough compared
to the shock standoff distance $r_{sh}$, 
and that the relativistic particles encounter a full shock compression
ratio while crossing the shock.
At the shock, density of matter will shoot up and inflow velocity will drop
abruptly. If $({\rho}_{-}, u_{-})$ and $({\rho}_{+}, u_{+})$ are the pre-
and post-shock densities and velocities respectively at the shock surface,
then
$$
\frac{{\rho}_{+}}{{\rho}_{-}} = R_{comp} = \frac{u_{-}}{u_{+}}
\eqno {(8)}
$$
where $R_{comp}$ is the shock compression ratio.
For high shock Mach number solution,
the expression for $R_{comp}$
can be well approximated as
$$
R_{comp} = 1.44{M_{sh}}^{\frac{3}{4}}\\
\eqno{(9)}
$$
where $M_{sh}$ is the shock Mach number and Eq. (5) holds for
$M_{sh} \gsim 4.0$ (Ellison \& Eichler 1985).\\
In terms of various accretion parameters, 
shock location can be computed as (Das 1999):
$$
r_{sh} = \frac{3{\sigma_{pp}}{\dot M}_{Edd}}{4{\pi}{u_{sh}}^2}\left(
\frac {1~-~2.4{M_{sh}}^{-0.68}}{1~-~3.2{{M_{sh}}^{-0.62}}}\right)
\eqno{(10)}
$$
where ${\sigma}_{pp}$ is the collision cross section for relativistic
protons, $u_{sh}$ and $M_{sh}$ are the dynamical flow velocity and the Mach 
number attained at the shock location, ${\dot M}_{Edd}$ is the mass
accretion rate scaled in units of Eddington rate. One can understand that \\
$$
\left(u_{sh}, M_{sh}\right)\equiv{\bf \zeta}\left({\cal E},{\dot M}_{Edd},
\gamma_{in}\right)
\eqno{(10a)}
$$
where ${\bf {\zeta}}$ has some complicated non-linear functional form
which cannot be evaluated analytically, but the value of $u_{sh}$ and
$M_{sh}$ can easily be obtained in terms of $\left\{{\cal E},{\dot M}_{Edd},
\gamma_{in}\right\}$ by numerically solving Eq. 5(a-b) and Eq. 10 (with the
help of Eq. 7(a-d)) simultaneously. Hence one can write \\
$$
r_{sh}\equiv{\bf {\xi}}\left({\cal E},{\dot M}_{Edd},\gamma_{in}\right)
\eqno{(10b)}
$$
where $\xi$ has some functional form other than that of $\zeta$.\\
\noindent
In ordinary stellar mass-loss computations (Tarafder 1988, and references therein),
the outflow is assumed
to be isothermal till the sonic point. This assumption is probably justified,
since
copious photons from the stellar atmosphere deposit momenta on the slowly outgoing
and expanding outflow and possibly make the flow close to isothermal. This
 need
not be the case for outflow from black hole candidates. Our effective boundary layer,
being close to the black hole, are very hot
and most of the photons emitted
may be swallowed by the black hole itself instead of coming out of the region
and
depositing momentum onto the outflow. Thus, the outflow could be cooler than
the isothermal flow in our case. We choose polytropic outflow with a different
polytropic index ${{\gamma}_o} < {\gamma_{in}}$ due to momentum deposition. 
Nevertheless, it may be advisable to study the isothermal outflow
to find the behaviour of the extreme case. Modelling the isothermal 
outflow is in progress and will be presented elsewhere. \\
\noindent
In our calculation we assume that essentially the post-shock 
fluid pressure and the post-shock proton temperature controls the
wind formation as well as the barionic matter content of the outflow.
The presence of a collision-less
steady standing spherical shock discussed in this work 
may randomize 
the directed in-fall motion
and at the shock surface the individual components of the total energy
of the flow (which is a combination of the kinetic, thermal and 
gravitational energy) gets rearranged in such a manner that the thermal energy
of the post-shock matter dominates (due to enormous shock generated post-shock proton 
temperature) over the gravitational attraction of the accretor
and a part of the in-falling material is driven by thermal pressure 
to escape to infinity as wind. In \S 3.3 we show that for any 
shock solution, whatever the initial boundary condition obeyed by the
pre-shock flow, the mass-loss rate always co-relates with 
the post-shock proton temperature which essentially supports the 
validity of our assumption.\\
\noindent
The adiabatic post-shock sound speed $a_{sh}^{+}$ and the 
post-shock temperature $T_{sh}^{+}$ (which is basically the 
temperature of the protons according to our one-temperature fluid 
approximation) can be calculated as:
$$
a_{sh}^{+}=\sqrt{\frac{{\gamma_o}p_{sh}^{+}}{\rho_{sh}^{+}}}
\eqno{(11a)}
$$
and
$$
T_{sh}^{+}=\frac{{\mu}{m_p}p_{sh}^{+}}{{\kappa}{\rho_{sh}^{+}}}
\eqno{(11b)}
$$
where $p_{sh}^{+}$ and $\rho_{sh}^{+}$ are the post-shock pressure and density 
of the flow at shock location $r_{sh}$ respectively. 
For low energy accretion (`cold' inflow, so to say) which is appropriate to produce 
a high shock Mach number solution, one can assume that the pre-shock thermal pressure
($p_{sh}^{-}$) may be neglected compared to its post-shock value
($p_{sh}^{+}$) and to the pre-shock ram pressure 
$\left({\rho_{sh}^{-}}\left(u_{sh}^{-}\right)^2\right)$.
One can obtain the value 
of $p_{sh}^{+}$
using Eq. (8-9) and from the total pressure 
balance condition at shock as,
$$
p_{sh}^{+} = \left({u_{sh}^{+}}\right)^2r_{sh}\left({\frac{R_{comp} - 1}{R_{comp}}}\right)
\eqno{(12)}
$$
Combining Eq. (8-12), post-shock sound velocity and temperature 
obtained at the shock surface can be rewritten as:
$$
a_{sh}^{+}=0.606{{\dot M}_{Edd}}{{\sigma_{pp}}^{1.5}}{\gamma_o}^{0.5}
\left({u_{sh}^{+}}\right)^{-1.5}\left(1-0.694{M_{sh}}^{-0.75}\right)^{0.5}
\left(\frac {1~-~2.4{M_{sh}}^{-0.68}}{1~-~3.2{{M_{sh}}^{-0.62}}}\right)^{1.5}
\eqno{(13a)}
$$
and,
$$
T_{sh}^{+}=0.014
\left(
\frac
{{\mu}m_p{\dot M}_{Edd}{\sigma_{pp}^{3}\left(u_{sh}^{+}\right)^{-3}}}{\kappa}\right)
\left(1-0.694{M_{sh}}^{-0.75}\right)
\left(\frac {1~-~2.4{M_{sh}}^{-0.68}}{1~-~3.2{{M_{sh}}^{-0.62}}}\right)^{3}
\eqno{(13b)}
$$
The two conservation equations governing the outflow will exactly be the same
as Eq. (5a) and (5b) with different polytropic index, i.e., 
with $\gamma=\gamma_o$ for outflow. So we can write,
$$
{\cal E}^{'}=h^{'}u^{'}_t
\eqno{(14a)}
$$
and
$$
{\dot {\cal M}}^{'}=4{\pi}
\left[\frac{\left(\gamma_o-1\right)a^o_2}{\gamma_o-\left(a^2_o+1\right)}\right]
^{\left({\frac{1}{\gamma_o-1}}\right)}
u_ou_tr^2
\eqno{(14b)}
$$
where ${\cal E}^{'}$
is the specific energy (including rest mass) 
of the outflow which is also assumed to be 
constant throughout the flow and
${\dot {\cal M}}^{'}={\dot M}_{out}{\gamma_{o}}^{\frac{1}{\gamma_{o}-1}}
{K^{o}}^{\frac{1}{\gamma_{o}-1}}$ is the entropy accretion rate of the outflow.
$\gamma_o~<~\gamma_{in}$ as already mentioned.\\
\noindent
Also Eq. (4a) can now be written as the mass conservation equation
for the outflow as:
$$
{\dot M}_{out}=4{\pi}{\rho^o}u^o\left({r}\right)^2
\eqno{(15)}
$$
In Eq. (15), $\rho^o$ and $u^o$ are the density and dynamical flow velocity of the
outflowing material.
We then define the mass outflow rate $R_{\dot m}$ as:
$$
R_{\dot m}=\frac{{\dot M}_{out}}{{\dot M}_{in}}
\eqno{(16)}
$$
\noindent
Like Eq. (7a-7d), the value of $\frac{du^o}{dr}$,
sonic point conditions and the value of $\frac{du^o}{dr}$
at the outflow sonic point can be written as:
$$
\left(\frac{du^o}{dr}\right)=\frac{u^o\left[1-\left({u^o}\right)^2\right]
\left[\left({a^o}\right)^2
\left(4r-3\right)-1\right]}
{2r\left(r-1\right)\left[\left({u^o}\right)^2-\left({a^o}\right)^2\right]}
\eqno{(17a)}
$$
$$
u_c^o=a_c^o=\sqrt{\frac{1}{4r_c^o-3}}
\eqno{(17b)}
$$
$$
\left(4r_c^o\right)^3\left[\left({{\cal E}^{'}}\right)^2-1\right]+
\left(4{r_c^o}\right)^2\left[2\left({{\cal E}^{'}}\right)^2{\varphi_{o}^2}-9\right]
+4r_c^o\left[\left({{\cal E}^{'}}\right)^2{\varphi_{o}^2}-27\right]+27=0
\eqno{(17c)}
$$
where ${\varphi_{o}}=\left(\frac{3\gamma_{o}+2}{\gamma_{o}-1}\right)$. \\
$$
\left(\frac{du^o}{dr}\right)_c^2+\frac{\left(\gamma_o-1\right)
\left[16\left(r^o_c\right)^2-16r_c^o-8{\gamma_o}r_c^o+6\gamma_o+3\right]}
{3r_c^o\left(4r_c^o-3\right)^{\frac{3}{2}}}
\left(\frac{du^o}{dr}\right)_c
$$
$$
+
\frac{\left(\gamma_o-1\right)\left(2r_c^o-1\right)\left[24\left(r^o_c\right)^2
-28r_c^o-
8\left(r^o_c\right)^2\gamma_o
+4r_c^o\gamma_o+3\gamma_o+6\right]}
{2r_c^o\left(4r_c^o-3\right)^{\frac{5}{2}}\left(r_c^o-
1\right)}
=0
\eqno{(17d)}
$$
Where the superscript $o$ indicates that the 
quantities are measured for the outflows.
\subsection{Simultaneous solution of inflow-outflow equations}
\noindent
In this work, we are interested in finding
the ratio of ${\dot M}_{out}$ to ${\dot M}_{in}$ (Eq. (16)), and
not the  explicit value of ${\dot M}_{out}$.
Also note that the primary goal of our present work was to compute the
outflow rate and to investigate its dependence on various inflow parameters but
not to study the collimation procedure of the outflow.\\
Before we proceed in detail, a general understanding of the transonic
inflow outflow system in the present case is essential to understand the basic scheme
 of the
solution procedure. Let us consider the transonic accretion first. Infalling matter
becomes supersonic after crossing a saddle-type sonic point, the location of which is
determined by $\left\{{\cal E},~{\dot M}_{Edd},~\gamma_{in}\right\}$. 
This supersonic
flow then
encounters a shock (if present), location of which ($r_{sh}$) is determined from
Eq. (10).
At the shock surface, part of the incoming matter, having a higher entropy density
(because shock in a fluid flow generates entropy),
is likely
to return as wind through a sonic point {\it other than} the point
through which it just entered.
Thus a combination of transonic topologies,
one for the inflow and other for the outflow
(passing through a {\it different}
 sonic point and following topology completely
different that of the `self-wind' of the accretion), is required to obtain a full
solution. So it turns out that finding a complete set of self-consistent
inflow outflow solutions ultimately boils down to locate
the sonic point of the polytropic outflow and the mass
flux through it. Thus a supply of parameters ${\cal E}, {{\dot M}_{Edd}}$,
${\gamma_{in}}$ and $\gamma_o$
make a self-consistent computation of ${R_{\dot m}}$ possible.
Here $\gamma_o$ is supplied as free parameter because the self-consistent
computation
of $\gamma_o$ directly using ${\cal E}, {{\dot M}_{Edd}}$ and
$\gamma_{in}$ has not been attempted in this work;
instead we put a constrain that $\gamma_o < \gamma_{in}$ always and for any value of
$\gamma_{in}$.
In reality, $\gamma_o$ is directly related to the heating and cooling 
processes taking place 
in the outflow.\\
\noindent
We obtain the inflow sonic point $r_c$ by solving Eq. (7c).
Using the fourth order
the Runge Kutta method, $u(r)$, $a(r)$ and
the inflow Mach number
$\left[\frac{u(r)}{a(r)}\right]$ are computed along the  inflow from the
{\it inflow} sonic point $r_c$ till
the position where the shock forms. The shock location is calculated
by solving Eq. (10).
Various shock parameters
(i.e., density, pressure etc at the shock surface) are then computed
self-consistently.\\
For outflow, with the known value of ${\cal E}^{\prime}$
and $\gamma_o$, it is easy to compute the location of the outflow sonic point 
$r_c^o$ 
from Eq. (17c). At the
outflow sonic point, the outflow velocity $u^o_c$ and polytropic sound 
velocity $a_c^o$ is computed
from Eq. (17b). Using Eq. (17a) and (17d),
$\left(\frac{du^o}{dr}\right)$
and $\left(\frac{du^o}{dr}\right)_c$ is computed as was
done for the inflow. Runge
-Kutta method is then
employed to integrate from the {\it outflow} sonic point $r_c^o$ towards the black hole
to
find out the outflow velocity $u^o$ and density $\rho^o$ at the shock location.
The mass outflow rate $R_{\dot M}$ is then computed using Eq.(16).\\
\noindent
It is obvious from the above discussion that $R_{\dot m}$ should have some
complicated non-linear functional dependence on the following accretion and
shock parameters:
$$
{R}_{\dot m} = {\bf {\Psi}}\left({\cal E},{{\dot M}_{Edd}},{r_{sh}},M_{sh},
{R_{comp}},{\gamma_{in}},{{\gamma}_o}\right)
\eqno{(18a)}
$$
As $r_{sh}, M_{sh}$ and $R_{comp}$ can be found in terms of ${\cal E},
{\dot M}_{Edd}$ and $\gamma_{in}$ only, ultimately it turns out that:
$$
{R}_{\dot m} = {\bf \Omega}\left({\cal E}, {\dot M}_{Edd}, \gamma_{in}, \gamma_o\right)
\eqno{(18b)}
$$
Where ${\bf {\Omega}}$ has some complicated functional form which cannot be
evaluated analytically.
\begin{figure}
\vbox{
\vskip -1.0cm
\centerline{
\psfig{file=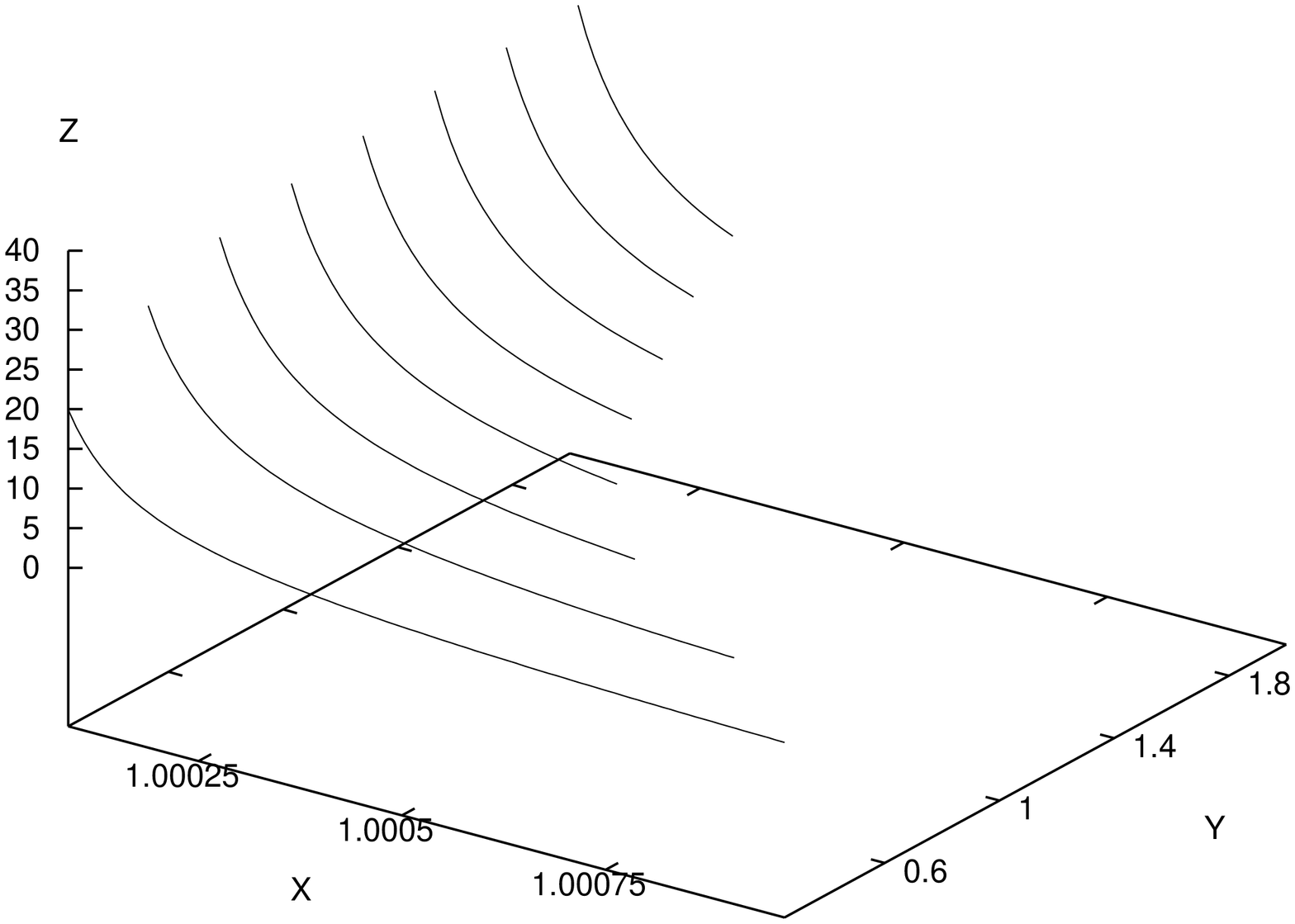,height=11cm,width=12cm,angle=270.0}}}
\noindent {{\bf Fig. 1:}
Variation of shock location $r_{sh}$
(in units of $r_g$) with conserved specific
energy ${\cal E}$ (which includes rest mass energy) and accretion rate
${\dot M}_{Edd}$ (scaled in units of Eddington rate) of the inflow for a
fixed value of inflow adiabatic index $\gamma_{in}$ (=$\frac{4}{3}$).
While ${\cal E}$ and ${\dot M}_{Edd}$ are plotted along X and Y axes
respectively, $r_{sh}$ is plotted along the Z axis.
It is observed that $r_{sh}$ correlates with ${\dot M}_{Edd}$ and
anti-correlates with ${\cal E}$.}
\end{figure}
\section{Results}
\subsection {Shock Location as a function of ${\cal E}$ and  
${\dot M}_{Edd}$}
\noindent
For a particular value of ${\cal E}$,
${\dot M}_{Edd}$ and $\gamma_{in}$,
the shock location (measured from the black hole in units of $r_g$) can be 
calculated using Eq. (10). As $u_{sh}$ and $M_{sh}$ is a function of {${\cal E}$,
${\dot M}_{Edd}$, and $\gamma_{in}$}, $r_{sh}$ will also change with the change of 
any of these accretion parameters. In Fig.1, we show the variation of $r_s$ 
with ${\cal E}$ and ${\dot M}_{Edd}$ for a fixed 
$\gamma_{in}=\frac{4}{3}$. 
While ${\cal E}$ and ${\dot M}_{Edd}$ are plotted along X and Y axes
respectively, $r_{sh}$ is plotted along 
Z axis in units of $r_g$. It is interesting to note that different curves 
start and terminate at different points, which indicates that shock formation 
is not a generic phenomena, i.e., shock does not form for any value of ${\cal 
E}$ and ${\dot M}_{Edd}$; rather, a specific region
of parameter space spanned by 
$\left\{{\cal E},{\dot M}_{Edd},\gamma_{in}\right\}$
allows shock formation. Also we find 
that depending on the value of ${\cal E}$, both sub-Eddington as well as 
super-Eddington accretion allows shock formation for a fixed value of 
$\gamma_{in}$. It is evident from the figure that while shock location 
non-linearly anti-correlates with ${\cal E}$ (for a given value of 
${\dot M}_{Edd}$ and $\gamma_{in}$), it correlates with ${\dot M}_{Edd}$ (for a 
fixed ${\cal E}$ and $\gamma_{in}$). 
Also we see that the maximum value of energy ${\cal E}_{Max}$ for which the 
shock forms decreases with an increase  of ${\dot M}_{Edd}$ and also the 
maximum as well as minimum value of ${\dot M}_{Edd}$ for which the shock 
forms, decreases and increases respectively with an increase in ${\cal E}$ for 
a fixed $\gamma_{in}$. Families of curves for other values of $\gamma_{in}$ can 
also be obtained in the same manner .
\begin{figure}
\vbox{
\vskip -2.5cm
\centerline{
\psfig{file=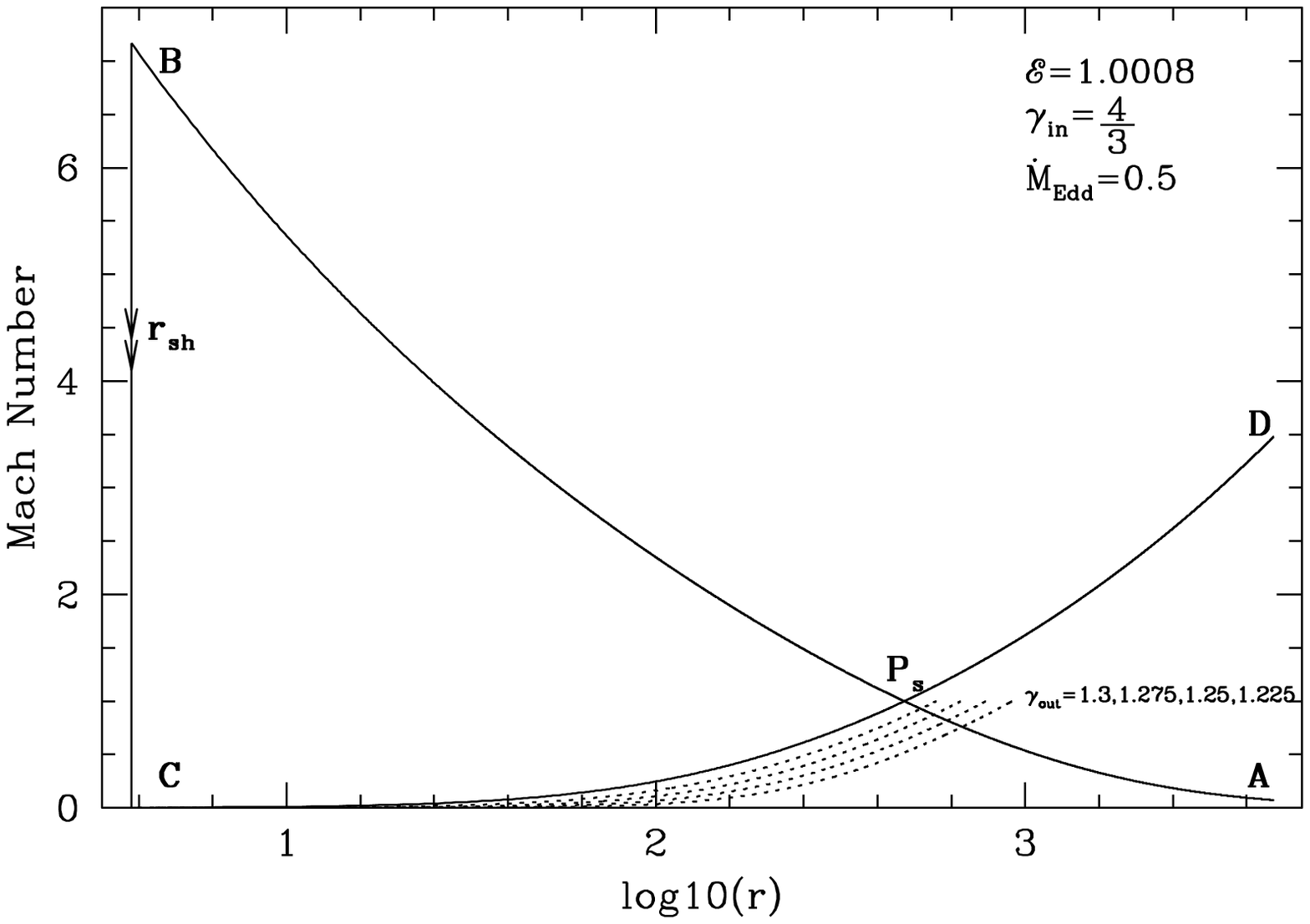,height=12cm,width=11cm}}}
\noindent {{\bf Fig. 2:}
Combined solution topology for transonic accretion-wind system
for fixed values
(shown in the figure)
of accretion parameters ${\cal E},~{\dot M}_{Edd},~
\gamma_{in}$ and for four outflow polytropic indices $\gamma_o=1.3,~1.275,~
1.25$ and $1.225$ respectively (from left to right). Solid curve marked with
AB represents the pre-shock transonic accretion while four dotted curves
represent the accretion-powered outflow branches. Solid vertical line BC
(marked by $r_{sh}$)
with double arrow stands for the shock transition and solid line marked by
CD stands for the `self-wind' branch. $P_s$ is the location of the
inflow sonic point. See text for details.}
\end{figure}
\subsection {Combined integral curves of motion}
\noindent
Fig. 2 shows a typical solution which combines the accretion and the
outflow.
While Mach number is plotted along the Y axis, the distance (in units of $r_g$)
from the
event horizon of the accreting black hole
is plotted along the X axis in logarithmic scale.
The accretion parameters used are ${\cal E}$ = 1.0008, ${\dot M}_{Edd}$
=0.5 
and $\gamma_{in} = \frac{4}{3}$
corresponding to ultra-relativistic inflow.
The solid curve AB represents the pre-shock region of the
inflow and the solid vertical line BC with double arrow at $r_{sh}$
represents
the shock transition. Shock location
(3.8 $r_g$) is obtained using the Eq.(10)
for a particular set of inflow parameters mentioned above.
Four dotted curves show the four different outflow branches corresponding 
to the four different adiabatic indices $\gamma_o$ of the outflow. From left 
to right, the values of $\gamma_o$ are 1.3, 1.275, 1.25 and 1.225 respectively 
with respective mass-outflow rates as  
0.0002, 0.00006, 0.000011 and 0.00000122 respectively which indicates 
that for a given value of ${\cal E}$, ${\dot M}_{Edd}$ and 
$\gamma_{in}$, $R_{in}$ correlates with $\gamma_o$,
 which is explicitly shown 
in Fig 6. It is evident from the figure that the outflow moves along the 
solution curves in a completely different way to that of the `self-wind' 
solution of the inflow (solid line marked by CD).
Also, the sonic points for 
all the outflowing branches are different to those of the accretion `self-wind' 
system which is designated as $P_s$. While $P_s=470.8~r_g$, the sonic points 
of the outflowing branches corresponding to $\gamma_o=1.3,1.275,1.25$ and $1.225$
are $574.96, 669.7, 783.34$ and $922.3~r_g$ respectively, which indicates that the 
outflow sonic point {\it increases} with a {\it decrease} in the adiabatic
index of the {\it outflow} and thus the wind starts with a very low bulk velocity which is 
why the mass-loss rate decreases. It is also observed that the sonic point of the 
accretion-`self-wind' system is, in general, located {\it closer} to the event
horizon compared to the outflow sonic point for {\it all values} of ${\cal E},
{\dot M}_{Edd}, \gamma_{in}$ and $\gamma_o$.
\begin{figure}
\vbox{
\vskip -3.0cm
\centerline{
\psfig{file=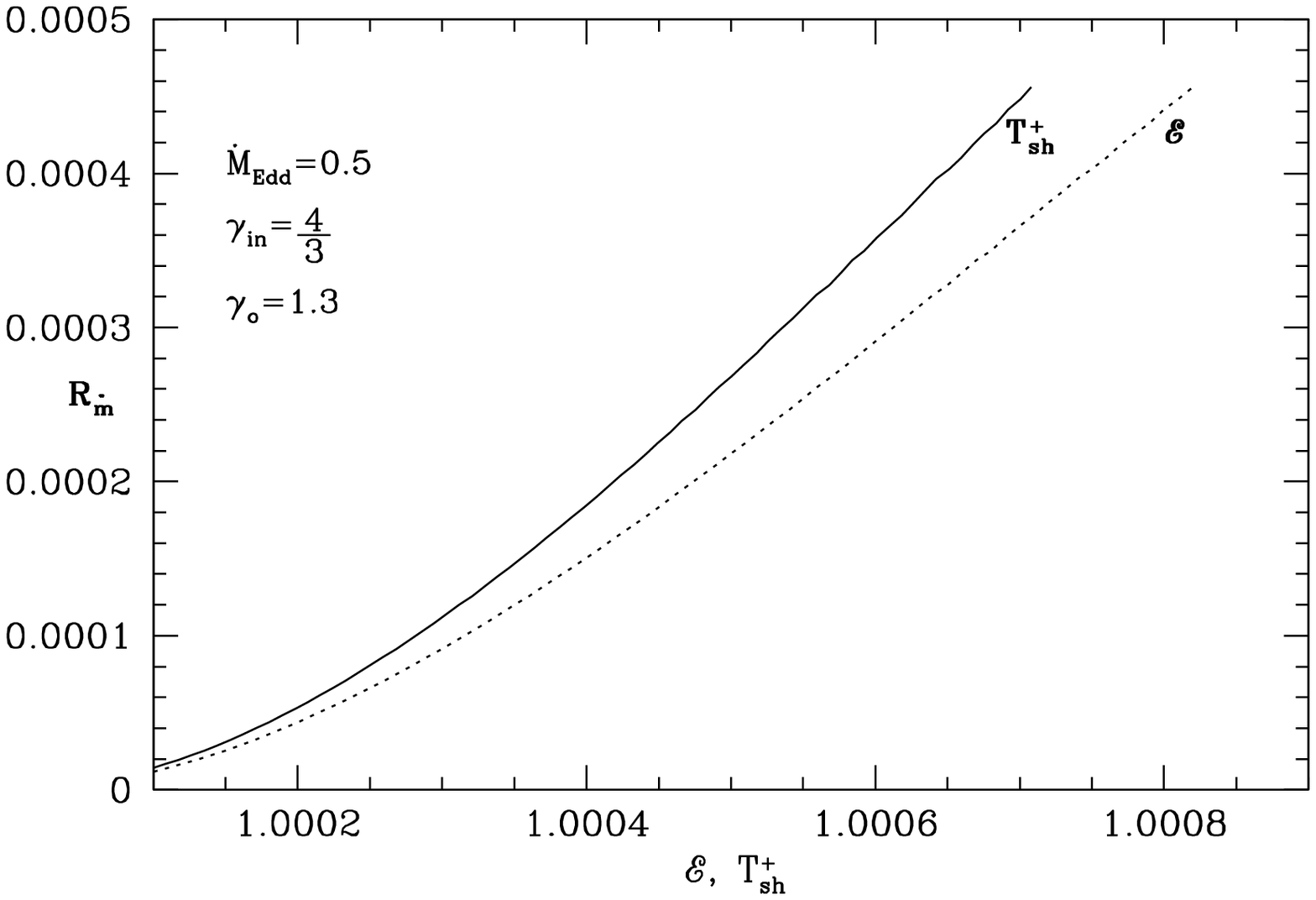,height=12cm,width=10.5cm}}}
\noindent {{\bf Fig. 3:}
Variation of $R_{\dot m}$ with conserved total specific energy
(which includes the rest mass energy) ${\cal E}$
(dotted curve in the figure marked with ${\cal E}$)
and post shock temperature (solid curve marked with $T_{sh}^{+}$) of
the flow for a fixed value of accretion rate and polytropic indices of the
flow having values shown in the figure. Both ${\cal E}$ and
$T_{sh}^{+}$ are plotted along X axis with the following scaling:\\
\noindent
$T_{sh}^{+} \longrightarrow~
\left[1+0.0004\left(\frac{T_{sh}^{+}}{T_{11}}\right)\right]$
(where $T_{11}=10^{11}~{^o\!K}$), to fit in the same graph. }
\end{figure}
\begin{figure}
\vbox{
\vskip -4.0cm
\centerline{
\psfig{file=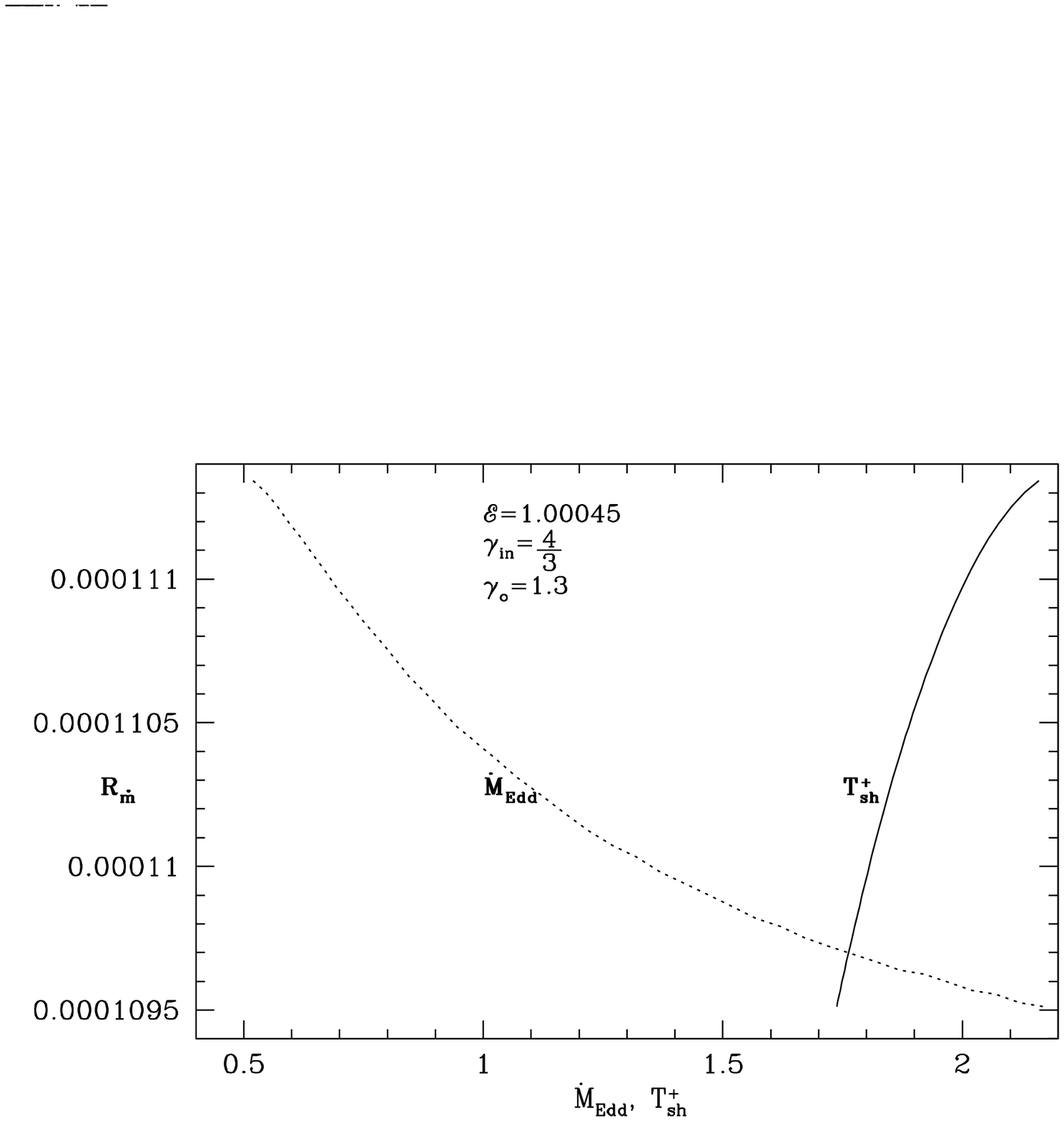,height=12cm,width=10.5cm}}}
\noindent {{\bf Fig. 4:}
Variation of $R_{\dot m}$ with accretion rate (dotted line
marked ${\dot M}_{Edd}$) and corresponding post-shock flow temperature
(solid line in the figure marked by $T_{sh}^{+}$) for  fixed values
of ${\cal E},~\gamma_{in},$ and $\gamma_o$ shown in the figure.
$T_{sh}^{+}$ is scaled as
$T_{sh}^{+}~\longrightarrow~2.2\left(\frac{T_{sh}^{+}}{T_{11}}\right)$.
It is evident from the figure that low-luminosity objects produce
more mass-loss, see text for details.}
\end{figure}
\subsection{Dependence of $R_{\dot m}$ on various flow parameters}
\subsubsection{Variation of $R_{\dot m}$ with ${\cal E}$}
\noindent
In Fig. 3, we have plotted the variation of $R_{\dot m}$ with conserved inflow
specific energy (which includes the rest mass energy of matter) ${\cal E}$
(dotted line marked with ${\cal E}$) for a fixed value of ${\dot M}_{Edd}~(=0.5)$,
$\gamma_{in}~\left(=\frac{4}{3}\right)$ and $\gamma_o~(=1.3)$. 
We observe that the mass-outflow rate non-linearly 
correlates with ${\cal E}$. The explanation may be as follows:\\
\noindent
As ${\cal E}$ increases, $r_{sh}$ decreases
(see Fig. 1, \S 3.1) and the post-shock bulk velocity of the flow 
$u_{sh}^{+}$ as well as the post-shock density $\rho_{sh}^{+}$
increases. 
The outflow rate, which is the product of three quantities $r_{sh}$,
$\rho_{sh}^{+}$ and $u_{sh}^{+}$ (see Eq. (15)),
increases in general due to the combined `tug of war'
of these three quantities.
Moreover, the closer the shock is to the black hole, the greater the
amount
of gravitational potential will be available to be put onto the relativistic
protons to provide stronger outward pressure  and the closer the shock forms
to the black hole, the higher is the post-shock temperature (the
effective characteristic outflow temperature) and the higher is the
amount of outflow (as the wind is observed to be strongly 
thermally driven, see discussion below).
Thus the mass-outflow rate increases with ${\cal E}$ because for a particular 
set of fixed values of ${\dot M}_{Edd}$, $R_{\dot m}$ is proportional to 
${\dot M}_{out}$ which increases with ${\cal E}$. Also in the same figure,
we show the variation of $R_{\dot m}$ with post-shock temperature 
$T_{sh}^{+}$ (solid line in the figure marked with $T_{sh}^{+}$). The value of 
$T_{sh}^{+}$ corresponds to the values of ${\cal E}$ shown in the same figure and is
scaled as 
$T_{sh}^{+}~\longrightarrow~
\left[1+0.0004\times \left(\frac{T_{sh}^{+}}{T_{11}}\right)\right]$
(where $T_{11}=10^{11}~{^o\!K}$) 
to fit in the same 
graph. We see that post-shock temperature correlates with the energy of the 
flow, and for a fixed accretion rate and adiabatic indices
of the inflow and outflow,
$R_{\dot m}$ also correlates with post-shock temperature, which indicates 
\begin{figure}
\vbox{
\vskip -4.0cm
\centerline{
\psfig{file=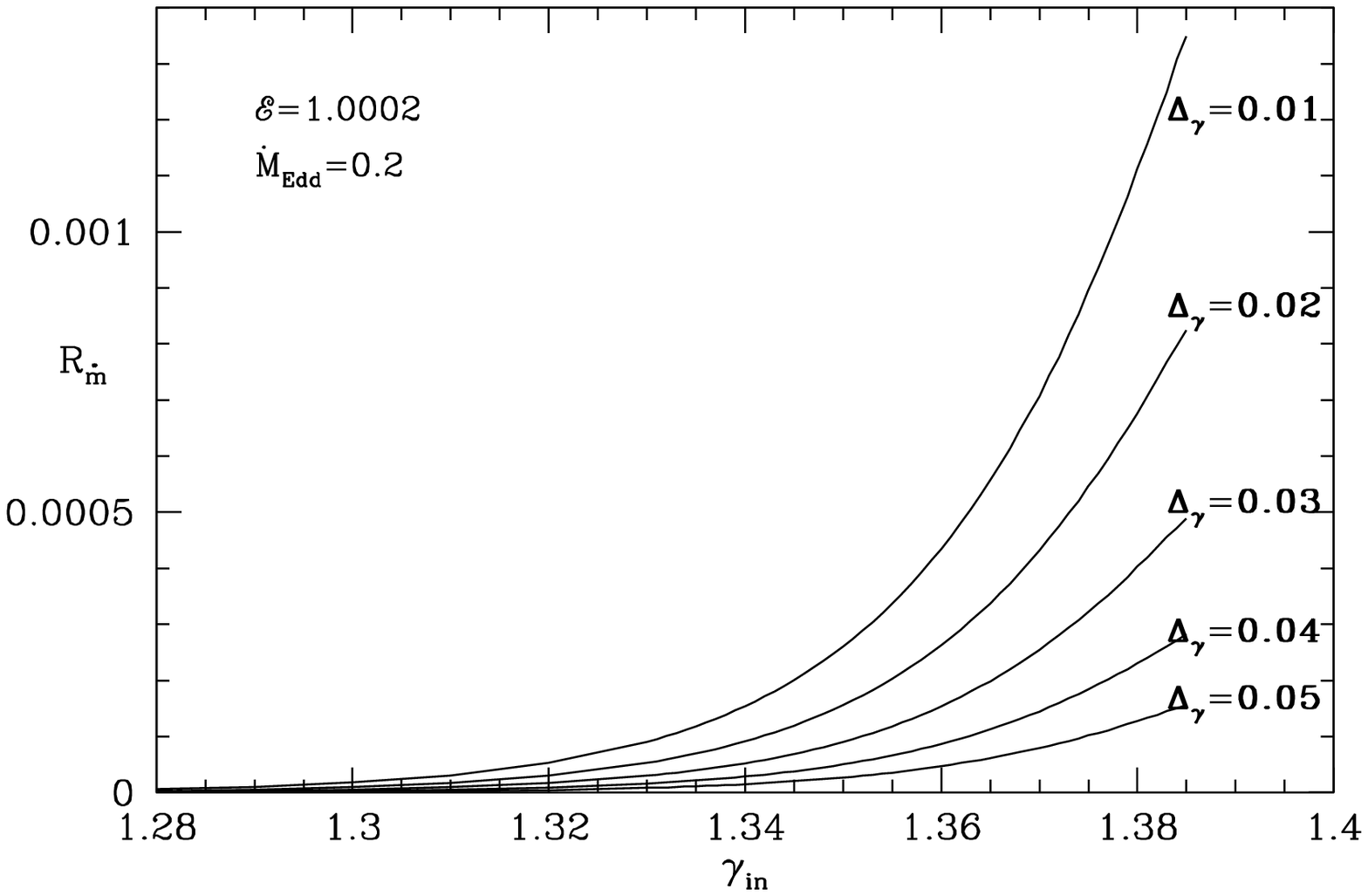,height=12cm,width=10.5cm}}}
\noindent {{\bf Fig. 5a:}
For fixed values of energy and accretion rate shown in the
figure, variation of $R_{\dot m}$ with polytropic index of the inflow
($\gamma_{in}$) is presented for five different values of
$\Delta_\gamma=\gamma_{in}-\gamma_o$, where $\gamma_o$ is the polytropic
index of the outflow. It is observed that ultra-relativistic
flows produce less mass-loss compared to pure-nonrelativistic flows, see
text for details.}
\end{figure}
\begin{figure}
\vbox{
\vskip -4.5cm
\centerline{
\psfig{file=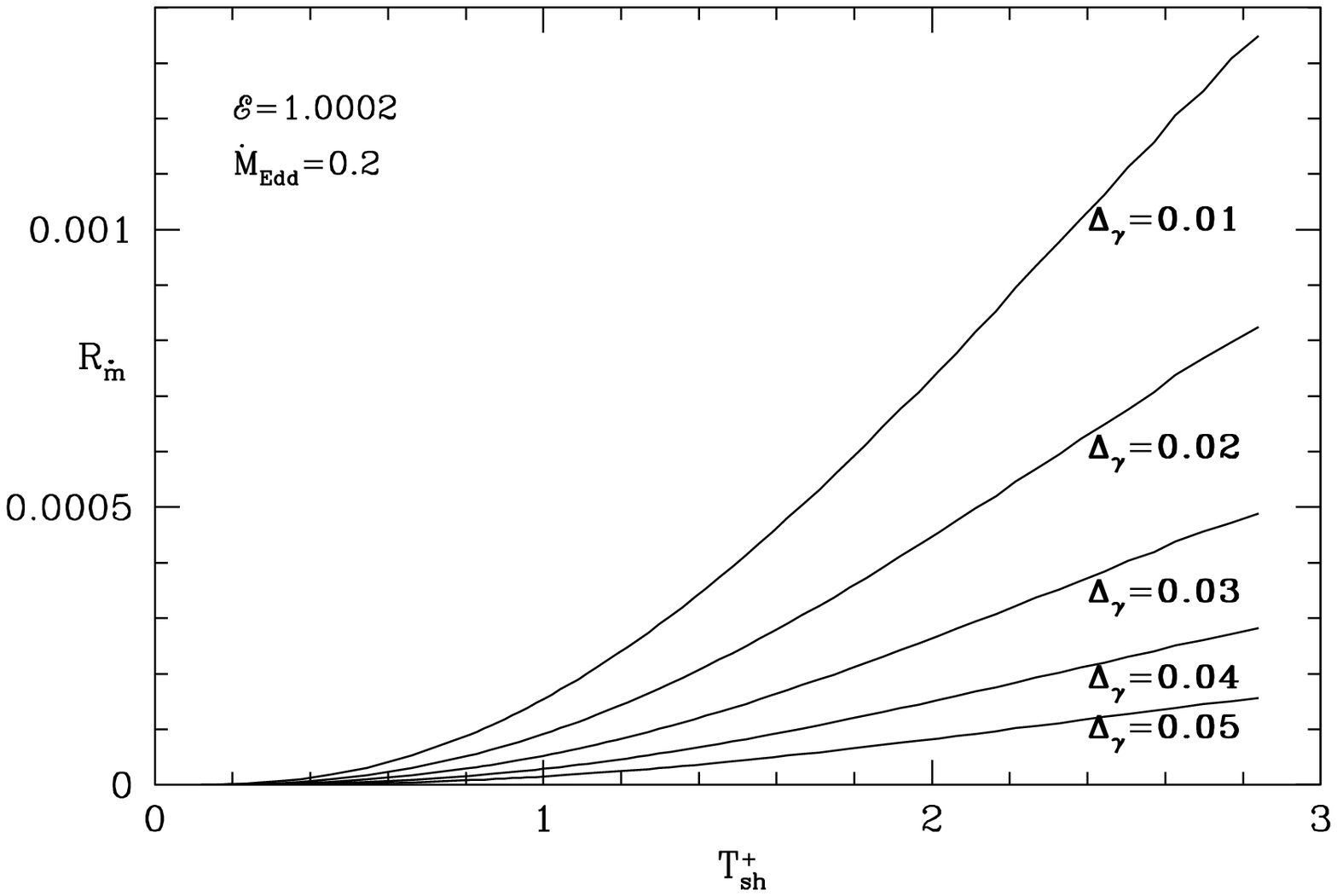,height=12cm,width=10.5cm}}}
\noindent {{\bf Fig. 5b:}
Variation of $R_{\dot m}$ with post-shock flow temperature
$T_{sh}^{+}$ corresponding to values of $\gamma_{in}$ shown in the figure
5(a). Outflow is observed to be thermally driven, see text for detail.}
\end{figure}
that outflow is thermally driven as well. If we draw a family of curves for 
$R_{\dot m}$ vs. ${\cal E}$ (alternatively, $R_{\dot m}$ vs. $T_{sh}^{+}$)
for different values of ${\dot M}_{Edd}$ (figure not presented in this paper)
ranging from sub-Eddington to super-Eddington accretion rate, we observe that
the maximum value of $R_{\dot m}$ obtained for a fixed ${\dot M}_{Edd}$ 
(while ${\cal E}$ is being varied) increases with 
a decrease of ${\dot M}_{Edd}$ and also the maximum value of $T_{sh}^{+}$
follows the 
same trend as mentioned above. So a `high energy-low luminosity' 
combination for polytropic accretion maximizes the post-shock flow 
temperature and gives rise to the highest amount of outflow.
\subsubsection{Variation of $R_{\dot m}$ with ${\dot M}_{Edd}$}
\noindent
In Fig.4 we show the variation of ${R_{\dot m}}$ as a function of accretion 
rate (dotted curve marked with ${\dot M}_{Edd}$)
scaled in units of Eddington 
rate for a fixed value of  ${\cal E}$ (=1.00045)
$\gamma_{in}~(=\frac{4}{3})$
and $\gamma_o$ (=1.3). It is observed that 
${R_{\dot m}}$ anticorrelates with ${\dot M}_{Edd}$ which indicates 
that
low-luminosity flow produces more mass-loss. Here ${R_{\dot m}}$ 
correlates with post shock temperature $T_{sh}^{+}$ (solid curve marked with 
$T_{sh}^{+}$). $T_{sh}^{+}$ shown here 
is scaled as $T_{sh}^{+}\longrightarrow~2.2\times \left(
\frac{T_{sh}^{+}}{T_{11}}\right)$ to fit in the same 
graph.The values of $T_{sh}^{+}$ correspond to the varying ${\dot M}_{Edd}$ 
shown in this graph.
It is also obvious from this figure that outflow is 
thermally driven because ${R_{\dot m}}$ correlates with $T_{sh}^{+}$. It can 
be shown that while for a fixed value of
${\cal E}$ and $\gamma_{in}$, $r_{sh}$,${\rho}_{sh}^{+}$ and $u_{sh}^{+}$ 
correlates with ${\dot M}_{Edd}$, shock Mack number, compression ratio and 
post-shock flow temperature anti-correlates with the accretion rate. 
However, from the figure we see that unlike ${\cal E}$, variation of
$R_{\dot m}$ 
is somewhat insensitive
to the variation of ${\dot M}_{Edd}$ for fixed 
${\cal E}$ and $\gamma_{in}$.
\begin{figure}
\vbox{
\vskip -4.0cm
\centerline{
\psfig{file=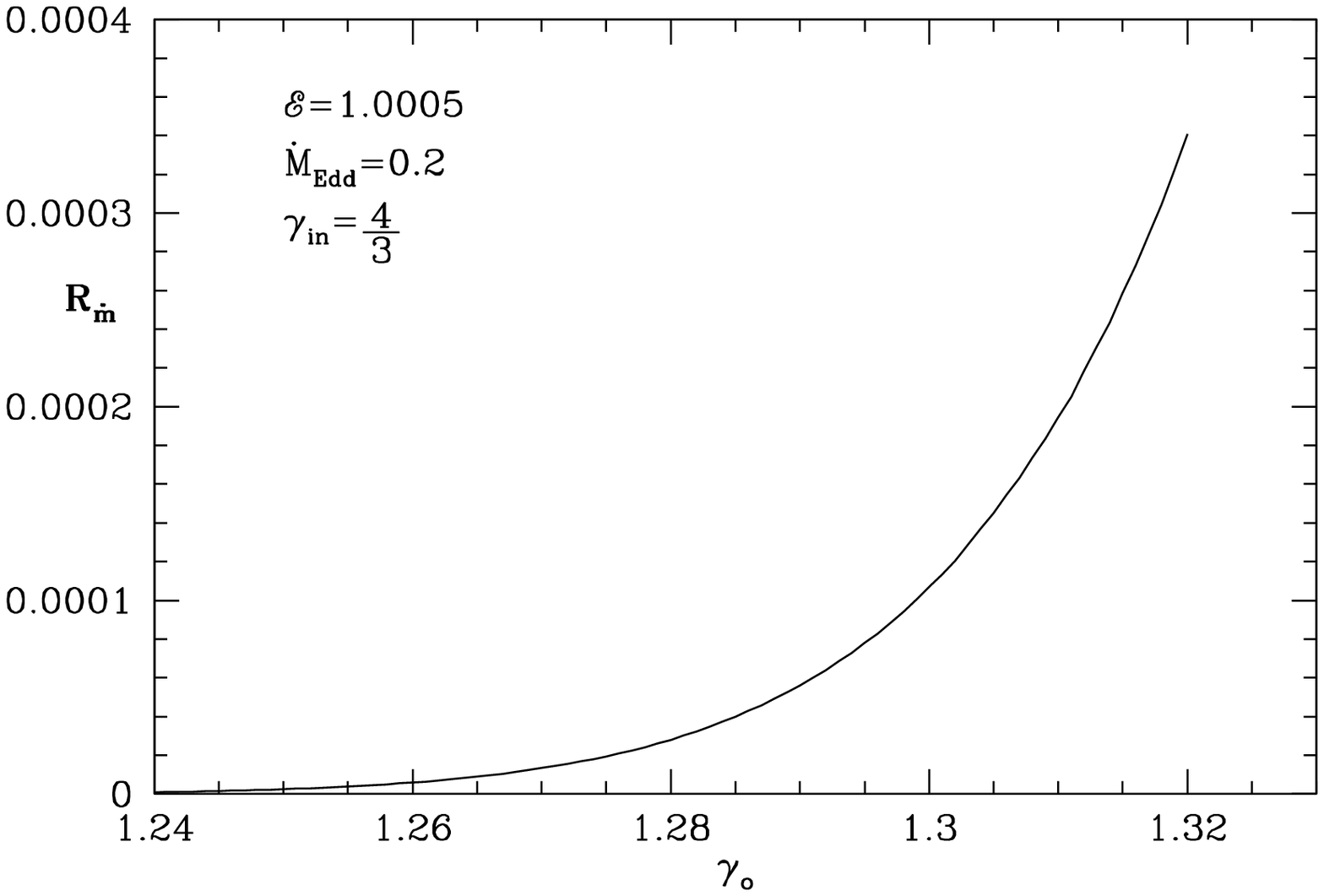,height=12cm,width=10.5cm}}}
\noindent {{\bf Fig. 6:}
Variation of $R_{\dot m}$ with polytropic index of the outflow
$\gamma_o$ for a fixed values of ${\cal E},~{\dot M}_{Edd}$ and
$\gamma_{in}$ as shown in the figure.}
\end{figure}
\subsubsection {Variation of ${R_{\dot m}}$ with adiabatic indices}
\noindent
In previous cases, the polytropic index $\gamma_{in}$ of the accreting matter
was always kept fixed at the value $\frac{4}{3}$. To have a better insight
of the behaviour of the outflow, we plot $R_{\dot m}$ as a function of
$\gamma_{in}$ (fig 5a) for a fixed value of ${\cal E}$ (= 1.0002)
 and
${\dot M}_{Edd}$ (= 0.2).
The upper 
range of $\gamma_{in}$ shown here is the range for which shock forms
for
the specified value of ${\cal E}$ and ${\dot M}_{Edd}$.
We have chosen the value of $\gamma_o$ in such a way so 
that the condition $\gamma_o<\gamma_{in}$ is always satisfied.
Defining $\Delta_\gamma$
to be 
${\Delta}_{\gamma}=\gamma_{in}-\gamma_o$,
we study the variation of ${R_{\dot m}}$ with $\gamma_{in}$ for five values of 
${\Delta}_{\gamma} = 0.01,~0.02,~0.03,~0.04$ and 0.05
(from 
top to bottom respectively). We observe that ${R_{\dot m}}$ correlates with 
$\gamma_{in}$, which is expected because the specific enthalpy of the flow 
increases with $\gamma_{in}$ to produce a higher post-shock temperature for 
higher value of $\gamma_{in}$ (see also Fig. 5(b)). We observe from our 
calculation that ${\rho}_{sh}^{+}$ and $u_{sh}$ correlates while $r_{sh}$ 
anticorrelates with $\gamma_{in}$. So increment of $\gamma_{in}$ satisfies 
all possible conditions to have a high value of $R_{\dot m}$.
In fig.5b, we 
show the variation of $R_{\dot m}$ with $T_{sh}^{+}$ (corresponding to the 
values of $\gamma_{in}$ shown in fig.5a)
scaled in units of 
$T_{11}=10^{10}~{^o\!K}$ to show that here also the flow is thermally driven. 
Also, we observe that $M_{sh}$ as well as the shock compression ratio 
$R_{comp}$           
anti correlates with $\gamma_{in}$ so that `weak-shock' 
solutions are preferred to obtain a high value of mass-loss for this case.
So we conclude that as the accretion approaches from its ultra-relativistic nature
to its non-relativistic regime, the mass-loss rate increases.\\
\noindent
In Fig. 6, we show the variation of $R_{\dot m}$ with $\gamma_o$ for a fixed 
value of ${\cal E}=1.0005,{\dot M}_{Edd}=0.2$ and $\gamma_{in}=\frac{4}{3}$. 
The general conclusion is that
$R_{\dot m}$ correlates with $\gamma_o$. This is because as $\gamma_o$
increases, shock location and post-shock density of matter does
not change (as $\gamma_o$ does not have any
role in shock formation or in determining the $R_{comp}$) but the sonic point
of the {\it outflow}
is pushed inward, hence  the velocity with which outflow leaves the shock surface
goes up, resulting the increment in $R_{\dot m}$.
\section {Conclusion}
\noindent
In this paper, we could successfully construct a self-consistent spherically-symmetric,
polytropic, transonic, non-magnetized inflow-outflow system by simultaneously solving the
set of hydrodynamic equations governing the accretion and wind around a Schwarzschild black hole
using full general relativistic framework. Introducing a steady, standing, hadronic-pressure
supported spherical shock (formation of which was first proposed by KE86 and PK83) 
surface around the black hole as the effective physical atmosphere which may be responsible
for generation of accretion-powered spherical wind, we calculate the mass-outflow
rate 
$R_{\dot m}$ in terms of only three accretion parameters (conserved energy of the flow 
${\cal E}$, which includes the rest mass energy of matter, accretion rate ${\dot M}_{Edd}$
scaled in units of Eddington rate and polytropic index of the flow $\gamma_{in}$) and
only one outflow parameter (the polytropic index of the outflow, $\gamma_o$).
Not only do we provide a sufficiently plausible
estimation of $R_{\dot m}$,
we could also successfully study
 the dependence and variation
of this rate on various physical parameters governing the flow.\\
\noindent
At this point, it is worth mentioning that the hot and dense
shock surface around black holes, which is proposed here as the effective
physical barrier around compact objects regarding the mass outflow, may
be generated due to other physical effects as well for spherical accretion
(Chang \& Osttriker 1985, M\'esz\'aros \& Ostriker 1983,
Babul, Ostriker \& M\'esz\'aros 1989, Park 1990, 1990a).
Another very important
approach launched recently was
to construct such an
`effective barrier' for non-spherical disc accretion
to introduce the concept of CENtrifugal pressure
supported BOundary Layers (CENBOL).
Treating the CENBOL as the effective atmosphere of the rotating flows around
compact objects
(which forms as a result of standing Rankine-Hugoniot shock or due to the
maximization of polytropic pressure of accreting material in absence of shock),
detailed
computation of the mass-outflow rate from the advective accretion
disks has been done, and dependence 
of this rate on various
accretion and shock parameters has been quantitatively studied
by constructing a
self-consistent disk-outflow system (Das 1998,
Das \& Chakrabarti 1999). \\
\noindent
The basic conclusions of this paper are the following:
\begin{enumerate}
\item Shock formation is not a generic phenomena, i.e., not all solutions contain
shock, rather a specific region of parameter space spanned by ${\cal E}, {\dot M}_{Edd}$
and $\gamma_{in}$ allows shock formation. Also we found that 
for given values of 
${\cal E}, {\dot M}_{Edd}$ and $\gamma_{in}$, while 
the value of shock location (in units of $r_g$)
{\it correlates} with 
${\dot M}_{Edd}$, it
{\it anti-correlates} with both ${\cal E}$ and $\gamma_{in}$.
\item The shock surface can serve as the `effective' physical barrier around the 
black hole regarding generation of mass loss via transonic spherical wind.
The fraction of accreting material being blown as wind (which is denoted as $R_{\dot m}$)
could be computed in terms of three accretion parameters and one outflow parameter.
\item While $R_{\dot m}$ {\it correlates} with ${\cal E}, \gamma_{in}$ and $\gamma_o$, 
it {\it anti-correlates} with ${\dot M}_{Edd}$, which indicates that low-luminosity objects
produce more mass-loss, though outflow could be generated for {\it both} sub-Eddington 
as well as super-Eddington accretion.
\item If a shock forms, then whatever the initial flow conditions and whatever the nature
of dependence of $R_{\dot m}$ on any of the accretion/ shock/ outflow parameters,
$R_{\dot m}$ {\it always correlates} with post-shock flow temperature, which indicates that
outflow is strongly thermally driven; hotter flow always produces more winds.
\end{enumerate}
\noindent
Our calculations in this paper, being
simply founded, do not explicitly include various radiation losses and cooling
processes, combined effects of which may reduce the post-shock proton temperature
(which means the reduction of outflow temperature), 
in reality could be lower than what we have obtained
here and the amount of outflow would be less than what is obtained in our calculation.
This deviation will be more important for systems
with high accretion rates.
Nevertheless, cases of low accretion rates discussed here would not be affected that
much and our preliminary investigation shows that even if we incorporate various losses,
the overall profile of the various curves showing the dependence of $R_{\dot m}$
on different inflow parameters would be exactly
the same, only the numerical value of $R_{\dot m}$
in some cases (especially for high accretion) might decrease. 
However, as we have shown that $R_{\dot m}$ anti-correlate with accretion rate 
( see \S 3.3.2, Fig. 4) and 
also it has been shown that the mass-loss rate does not have that strong a dependence on 
${\dot M}_{Edd}$, we believe that such changes will not
bring any significant change 
in our overall conclusions.\\
\noindent
Our present work, as we believe, may have
some important consequences regarding modelling of the
astrophysical outflows emanating from galactic and extra-galactic sources
powered by isolated accreting compact objects.
Galactic and extra-galactic sources of jets and outflows
are now widely believed to be fed by accreting black holes sitting at the dynamical 
centre
of these sources. In the absence of any binary companion, spherically symmetric
 Bondi (Bondi
1952) type accretion may occur onto isolated central black hole
if accreting matter has a negligible amount of intrinsic angular momentum.
This may happen when the central super-massive black hole
at the galactic centre is surrounded by a dense stellar cloud in such a way
that the vector sum of the angular momentum of tidally disrupted
matter (from a number of stars with trajectories approaching sufficiently
close to the hole) almost vanishes.
On the other hand, unlike the ordinary stellar bodies, black holes do not have their own
`physical' atmosphere and outflows in these cases have to be generated from the
accreting matter only. Hence we believe that it is {\it necessary} to study the
accretion and outflow (from various astrophysical sources powered by accreting compact
objects) in the same framework instead of treating the outflow separately from
accretion phenomenon. At the same time, as the fundamental criterion for
constructing
any self-consistent physical model demands the minimization of the number of inputs
to the model, we may conclude that modelling the astrophysical outflow needs a
concrete formulation where the outflow can be described in terms of {\it minimum}
number
of physical parameters governing the inflow. The success of our work presented in this
paper, as we believe, is precisely this. We could rigorously compute the mass 
outflow rate
$R_{\dot m}$ in terms of only three inflow parameters
and only one outflow
parameter.
To the best of our knowledge, there is no such model present in the
literature which rigorously studies the accretion-powered spherical
outflow 
in a general relativistic framework using only the four parameters mentioned above.\\
\noindent
Although in this work we have performed our calculation for a 10$M_{\odot}$ Schwarzschild
black hole, general flow characteristics will be unchanged for black hole of any
mass except the fact that the region of parameter space responsible for shock formation
will be shifted and the value of $R_{\dot m}$ will explicitly depend on the
mass of the black hole. In our next work, we will show
the direct dependence of $R_{\dot m}$ on the mass of the black hole and will apply our
model
to some specific astrophysical sources of outflows to roughly estimate the amount
of outflow in $M_{\odot}/{yr}^{-1}$. 

\begin{acknowledgements}
The author would like to acknowledge the hospitality provided by the Astronomy Unit,
School of Mathematical Sciences, Queen Mary \& Westfield College, University of London, 
where a part of this research was carried out.
\end{acknowledgements}
{}
\end{document}